\documentstyle[12pt]{article}
\input{amssym.def}
\input{amssym.tex}
\def\I{{\rm i\,}}
\def\bbbr{{\Bbb R}}
\def\bbbc{{\Bbb C}}
\def\D{{\rm d\,}}
\def\E{{\rm e\,}}

\begin{document}

\noindent {\Large {\bf $N$-order Darboux transformation
and a spectral problem on semiaxis }}

\vspace{3mm}
\begin{quotation}
\noindent
Vladislav G.\,Bagrov$^{a,b}$, Boris F.\,Samsonov$^a$, and L.
A.\,Shekoyan$^a$

\noindent
$^a$Tomsk State University, 36 Lenin Avenue, 634050 Tomsk, Russia

\noindent
$^b$High Current Electronics Institute, 4 Akademichesky Avenue,
634055 Tomsk, Russia

\vspace{3mm}\noindent
{\bf Abstract.}\
$N$-order Darboux transformation operator is defined on the basis of a general notion of transformation operators.
Factorisation properties of this operator are
studied. The Darboux transformation operator technique is applied to
construct and investigate potentials with bound states at arbitrary energies for the spectral problem on semiaxis.
\end{quotation}

\vspace{3mm}
\noindent {\bf 1. Introduction}

\vspace{2mm}
\noindent In the inverse quantum scattering theory one usually uses
the integral transformation operators (see, e.g., \cite{invscat}). An
essential element in their construction is the Gelfand-Levitan-Marchenko
equation. The differential transformation operators, such as the Darboux
transformation operators (\cite{dar}), have no such a prevalence. It was
established in recent papers (\cite{tmf,Jph}) that the integral
transformation with a degenerate kernel is equivalent to a differential one.
This possibility has been earlier noted by L. D. Faddeev \cite{Fad} and
was recently discussed in (\cite{SL}). Many properties of Darboux
transformations are studied in the book by V.B. Matveev and M.A. Salle
\cite{MatvSal}. They define this transformation on the basis of covariance property
of the Schr\"odinger equation with respect to a transformation of the wave
function and potential energy. We confine ourselves with other definition
\cite{tmf,Jph,BSrev}
Our definition is based on a general notion of transformation operators introduced and investigated by Delsart and Lions \cite{DelLion} (see also \cite{Levitan}). In terms of these notions Darboux \cite{dar} investigated differential first order transform
ation operators for the Sturm-Liouville problem. This is the reason, in our opinion, to call
any differential transformation operator {\it Darboux transformation operator}.

In recent years one has investigated such transformations in
connection with the spectral properties of the Schr\"odinger operator (\cite
{FA}) and as a source of new exactly solvable potentials (\cite{dub,BSrev})
and reflectionless potentials with an infinite discrete spectrum (\cite
{shabat}). In the supersymmetric quantum mechanics the $N$-order
differential transformation operators are used in the construction of higher
derivative supercharge operators (\cite{andr}) and lead to new peculiarities
in the supersymmetry breakdown (\cite{mpla}).
In this connection the investigation
of the properties of the $N$-order differential transformation operators is
of interest.

In this paper we define the $N$-order Darboux transformation
operator and give an improved formulation and a new proof of the theorem about the factorisation of this
operator by the first order Darboux transformation operators which represent
a dressing chain. It is established that the ordinary Darboux transformation
operator provide a one-to-one correspondence between the solution spaces of
the input and output Schr\"odinger equations. We then apply this technique
to construct and investigate potentials with given discrete spectrum for
 the spectral problem on semiaxis. We obtain
an exactly solvable equation with arbitrarily
disposed discrete spectrum levels. We give for the latter equation discrete
spectrum eigenfunctions, Jost solution, and Jost function. The expression
for the Jost function shows that the potential so obtained pertains to the
Bargman type potentials.
Other applications of these transformations may be found in \cite{tmf,Jph,BSrev},\cite{FA}-\cite{mpla}.

\vspace{3mm}\noindent {\bf 2. $N$-order Darboux transformation operator}

\vspace{3mm}\noindent Let us consider the one dimensional Schr\"odinger equation
\begin{equation}
\label{e1}h_0\psi (x)=E\psi (x),\quad h_0=-D^2+V_0(x),\quad D=-\D /\D %
x,\quad x\in [0,\infty )
\end{equation}
where $V_0(x)$ is a sufficiently smooth real valued function. Let $T_0$ be
the functional (if necessary topological) space of the solutions of the
equation (\ref{e1}).

{\sc Definition 1.} {\it Let us call the }$N${\it -order linearly
differential operator $L^{(N)}$ with the coefficient at $D^N$ equal to
unity, acting from $T_0$ to
$$
T_{N1}=\left\{ \varphi :\ \varphi =L^{(N)}\psi ,\ \psi \in T_0\right\} ,
$$
{\em an }}$N${\it {\em -order Darboux transformation operator} for the
Hamiltonian $h_0$ if it satisfies the following operator equation:
\begin{equation}
\label{e2}L^{(N)}h_0-h_0L^{(N)}=A_N(x)L^{(N)},
\end{equation}
where $A^{(N)}(x)$ is a sufficiently smooth function. If $A^{(N)}(x)\equiv 0$%
, the operator $L^{(N)}$ is called {\em trivial}.}

{\sc Remark 1}. {\it Every trivial transformation operator is a
symmetry operator for equation (\ref{e1})}.

It follows from Definition 1 that the function $\varphi =L^{(N)}\psi
$ satisfies the Schr\"odinger equation with the potential $%
V_N(x)=V_0(x)+A_N(x) $ (transformed Schr\"odinger equation) and the space $%
T_{N1}\subset T_N$, where $T_N$ is a functional space of the solutions of
the latter equation.

For $N=1$ equation (\ref{e2}) defines the well known Darboux
transformation with the transformation operator of the form
\begin{equation}
\label{e3}L^{(1)}=L=-u_\alpha ^{\prime }/u_\alpha +D,\quad A_1(x)=-2\left(
\log u_\alpha \right) ^{\prime \prime },
\end{equation}
where the prime denotes the derivative with respect to $x$. The function $%
u_\alpha $ called {\it a transformation function} is defined by the initial
Hamiltonian $h_0$: $h_0u_\alpha =\alpha u_\alpha $, $\alpha \in \bbbr$, $%
\mathop{\rm Im}u_\alpha =0$. It is clear from (\ref{e3}) that $\ker L=%
\mathop{\rm span}\left\{ u_\alpha \right\} $, where ''$\mathop{\rm span}$''
stands for the linear hull over the complex number field $\bbbc$.

If $\widetilde{u}_\alpha $ is chosen such that $W\left( u_\alpha ,%
\widetilde{u}_\alpha \right) =1$ where $W$ stands for the Wronskian of the
functions $u_\alpha $ and$\ \widetilde{u}_\alpha $, then we have $L%
\widetilde{u}_\alpha =u_\alpha ^{-1}=v_\alpha $ and $h_1v_\alpha =\alpha
v_\alpha $, $h_1=h_0+A_1(x)$. It is not difficult to convince ourselves that
$$
\lim \nolimits_{E\rightarrow \alpha }R^{-1}(E)L\psi _E(x)=\widetilde{v}%
_\alpha (x),\quad R(E)=E-\alpha
$$
under the condition $\psi _E(x)\rightarrow u_\alpha (x)$ as $E\rightarrow
\alpha $. In this case we have $h_1\widetilde{v}_\alpha =\alpha \widetilde{v}%
_\alpha $ and $W\left( v_\alpha ,\widetilde{v}_\alpha \right) =W\left(
u_\alpha ,\widetilde{u}_\alpha \right) =1$. Therefore we can always define
on the space $T_0$ a linear operator $\widehat{L}$ by putting%
$$
\begin{array}{c}
\widehat{L}\psi _E=R^{-1/2}\left( E\right) L\psi _E,\quad \forall E\neq
\alpha , \\ \widehat{L}\widetilde{u}_\alpha =L\widetilde{u}_\alpha =v_\alpha
=u_\alpha ^{-1}\quad \mbox{ and }\quad \widehat{L}u_\alpha =\widetilde{v}%
_\alpha \enspace .
\end{array}
$$
The operator $\widehat{L}$ maps every solution of the Schr\"odinger equation
with the Hamiltonian $h_0$ to the solution of the same equation with the
Hamiltonian $h_1$ and $W\left( \varphi _E,\widetilde{\varphi }_E\right)$ $
=W\left( \psi _E,\widetilde{\psi }_E\right) $, $\forall \psi _E,\widetilde{%
\psi }_E\in T_0$.

It follows from (\ref{e2}) that for the real valued function $A_N(x)$
the operator $L^{(N)^{+}}L^{(N)}$ where $L^{(N)^{+}}$, is an operator
formally conjugate to $L^{(N)}$ is a $2N$-order differential symmetry
operator for equation (\ref{e1}) and consequently it is an $N$-order
polynomial with respect to $h_0$.

We will first consider three lemmas which are necessary for the
proof of the main theorem.

{\sc Lemma 1.}{\it \ Operator $L\equiv L^{\left( 1\right) }$ is the
Darboux transformation operator if and only if $L^{+}L=h_0-\alpha $, $\alpha
\in \bbbr$.}

This lemma can easily be proved by direct calculations.

Since we have $\ker L^{+}=\mathop{\rm span}\left\{ v_\alpha
=u_\alpha ^{-1}\right\} $, we can define an operator $\widehat{L}^{+}$in the
space $T_1 $ by putting
$$
\widehat{L}^{+}\varphi _E=R^{-1/2}\left( E\right) L^{+}\varphi _E,\quad
\forall E\neq \alpha
$$
and
$$
\widehat{L}^{+}\widetilde{v}_\alpha =L^{+}\widetilde{v}_\alpha =u_\alpha
=v_\alpha ^{-1},\quad \widehat{L}^{+}v_\alpha
=\widetilde{u}_\alpha \enspace %
.
$$
The operators $\widehat{L}$ and $\widehat{L}^{+}$ assure a one-to-one
correspondence between the spaces $T_0$ and $T_1$. Moreover, we have
$$
\begin{array}{c}
T_0=T_{01}\cup
\mathop{\rm span}\left\{ \widetilde{u}_\alpha \right\} ,\quad \
T_1=T_{11}\cup \mathop{\rm span}\left\{ \widetilde{v}_\alpha \right\} , \\ \
T_{01}=\{\psi :\ \psi =L^{+}\varphi ,\ \varphi \in T_1\}\enspace .
\end{array}
$$

{\sc Lemma 2.} {\it Operator $L\equiv L^{(2)}$ can always be
presented in the form $L=L_2L_1$, where $L_1=-u_1^{\prime }/u_1+D$ and $%
L_2=-v^{\prime }/v+D$ are the first order Darboux transformation operators, $%
u_1$ is the transformation function satisfying equation (\ref{e1}) with the
eigenvalue $C_1$ $v$ is the transformation function for the iterated Darboux
transformation satisfying the Schr\"odinger equation with an intermediate
potential $V_1$ obtained after the Darboux transformation with the operator $%
L_1$ and corresponding to the eigenvalue $C_2$. If $C_1$ and $C_2\in \bbbr $%
, then they are arbitrary and the functions $u_1$ and $v$ are real valued.
If $C_1$ and $C_2\in \bbbc $, then $C_2=\overline{C}_1$ (overline signifies
the complex conjugation) and $v=L_1\overline{u}_1$. The potential difference
is a real valued function.}

We note first of all that a similar statement is discussed in the
paper by Andrianov et al. (1995) but we need here some details of the proof
and so we give a complete proof of this lemma.

Consider a second order differential operator of the general form $%
L=a_0(x)+a_1(x)D+a_2(x)D^2$. Equation (\ref{e2}) reduces to a system of
differential equations for the coefficients $a_i(x)$, $i=0,1,2$ and $%
A(x)\equiv A_2(x)$. It follows from this system that $a_2=\mathop{\rm const}$
and without the loss of generality we put $a_2=1$. We find then $%
A=2a_1^{\prime }$. After excluding $a_0$ and $A$ we obtain a differential
equation for the function $a_1$ which can readily be twice integrated with
integration constants $2\alpha _1$, $\alpha _2\in \bbbr$. As a result we
obtain the following differential equation for the function $a_1$:
\begin{equation}
\label{e4}a_1^2V_0+a_1^2a_1^{\prime }-\frac 12a_1a_1^{\prime \prime }+\frac
14a_1^{\prime 2}-\frac 14a_1^4-\alpha _1a_1^2-\alpha _2=0\enspace .
\end{equation}
After introducing a new dependent variable $u_1$ with the help of the
following relation
\begin{equation}
\label{e5}u_1^{\prime }/u_1=\frac 12a_1^{\prime }/a_1-\frac 12a_1-\sqrt{%
\alpha _2}/a_1,
\end{equation}
we rewrite equation (\ref{e4}) in the form
$$
-D^2u_1+(V_0-C_1)u_1=0,
$$
where $C_1=\alpha _1-\sqrt{\alpha _2}$. We state consequently that the
function $u_1$ is a solution of the initial Schr\"odinger equation. Since
the solutions of this equation are supposed to be known, the function $u_1$
is determined. We solve now equation (\ref{e5}). For this purpose we
introduce a new function $v$ by putting $a_1=-[\ln (vu_1)]^{\prime }$.
Equation (\ref{e5}) reduces to the following equation for the function $v$:
\begin{equation}
\label{e6}-D^2v+(V_1-C_2)v=0,
\end{equation}
where $C_2=\alpha _1+\sqrt{\alpha _2}$ and $V_1=V_0-2(\ln u_1)^{\prime
\prime }$. Equation (\ref{e6}) shows that the function $v$ is a solution of
the intermediate Schr\"odinger equation obtained by the Darboux
transformation with the operator $L_1$ and the transformation function $u_1$%
. Taking into account the expressions for the functions $a_1$ and $a_0$
$$
a_1=-[\ln (vu_1)]^{\prime },\quad a_0=u_1^{\prime }v^{\prime }/(u_1v)-(\ln
u_1)^{\prime \prime },
$$
we obtain an expression for the operator $L$ formulated in the assertion of
the lemma. For $C_2\ne C_1$ the function $v$ is a transform of the function $%
u_2$: $v=L_1u_2=u_1^{-1}W(u_1,u_2)$ which is a proper function of the
initial Hamiltonian: $h_0u_2=C_2u_2$. The potential difference is given by
the expression
\begin{equation}
\label{e7}A_2=-2[\ln W(u_1,u_2)]^{\prime \prime }\enspace .
\end{equation}
For $C_2=C_1$ we have $v=\beta _1u_1^{-1}+\beta _2\widetilde{v}$, where $%
\beta _1$, $\beta _2\in \bbbr$ and $u_1^{-1}$, $\widetilde{v}$ are linearly
independent solutions of equation (\ref{e6}).The Wronskian $W$ should be
replaced in this case by $\beta _1+\beta _2u_1\widetilde{v}$. 
$\square $

{\sc Corollary 1.}{\it \ It follows from Lemmas 1 and 2 that }
$$
L^{+}L=(h_0-C_1)(h_0-C_2),
$$
$$
LL^{+}=(h_2-C_1)(h_2-C_2),\quad h_2=h_0+A_2\enspace .
$$

{\sc Remark 2.}{\it \ For }$C_1=C_2=C\in \bbbr${\it \ and }$%
v=u_1^{-1}${\it \ (}$\beta _2=0${\it ) the operator }$L=-L_1^{+}L_1=C-h_0$%
{\it \ is a trivial transformation operator.}

{\sc Remark 3. }{\it In the case $C_2\ne C_1$, the intermediate
transformation function can be expressed through the solution $u_2$ of the
initial equation (\ref{e1}). In this case we obtain the known (\cite
{CrumKrein}) expression for the transformation operator, which in the
general case has the form }
\begin{equation}
\label{LN}L^{(N)}=L_NL_{N-1}...L_1=W^{-1}(u_1,...,u_N)\left|
\begin{array}{cccc}
u_1 & u_2 & \ldots & 1 \\
u_1^{\prime } & u_2^{\prime } & \ldots & D \\
\cdots & \cdots & \cdots & \cdots \\
u_1^{(N)} & u_2^{(N)} & \ldots & D^N
\end{array}
\right| \enspace .
\end{equation}
{\it In addition, the following factorizations are valid (\cite{andr,tmf}): }%
$$
L^{(N)^{+}}L^{(N)}=P(h_0)=\prod_{i=1}^N(h_0-C_i){\it ,\quad }%
L^{(N)}L^{(N)^{+}}=P(h_N)
$$
{\it where }$h_N=h_0+A_N${\it , }$A_N=-2\left[ \log W(u_1,...,u_N)\right]
^{\prime \prime }${\it , }$h_0u_i=C_iu_i${\it \ and all }$C_i${\it \ are
different. If the coefficients of the polynomial $P\left( x\right) $ belong
to the field $\bbbr$, then the intermediate potentials (in the case where
the polynomial $P\left( x\right) $ has complex zeros) can be complex valued,
but the final potential is real valued if the transformation functions are
chosen appropriately.}

Let us denote by $C_1,\ldots ,C_N$ the zeros of the polynomial $%
P\left( x\right) $ which can be of an arbitrary order.

{\sc Lemma 3.}{\it \ If $L^{(N)}$ is an $N$-order Darboux
transformation operator with the transformation functions $u_i$ such that $%
h_0u_i=C_iu_i$, then we have
$$
\ker L^{(N)}\cap \bigcup_{i=1}^N\ker (h_0-C_i)\neq \emptyset \enspace .
$$
}

Consider one of the zeros of the polynomial $P\left( x\right) $,
e.g., $C_1$. If $\ker L^{(N)}\cap \ker (h_0-C_1)\neq \emptyset $, then the
lemma is proved. Let we have
\begin{equation}
\label{e8}\ker L^{(N)}\cap \ker (h_0-C_1)=\emptyset \enspace .
\end{equation}
Consider $v_1=L^{(N)}u_1$ and $\widetilde{v}_1=L^{(N)}\widetilde{u}_1$,
where $u_1$ and $\widetilde{u}_1$ form a basis in the space $\ker (h_0-C_1)$%
. By virtue of the linearity of the operator $L^{\left( N\right) }$ and
assumption (\ref{e8}), the space $\mathop{\rm span}\left\{ v_1,\widetilde{v}%
_1\right\} $ cannot be a one dimensional space. Then we have
\begin{equation}
\label{e9}\mathop{\rm span}\left\{ v_1,\widetilde{v}_1\right\} =\ker
(h_N-C_1)\subset \ker L^{(N)^{+}}\enspace .
\end{equation}
The operator $L^{(N)^{+}}$, with the help of Proposition 2.1 given in the
book by Berkovich \cite{berk}, can be presented in the form $%
L^{(N)^{+}}=L^{(N-2)^{+}}L_2^{+}L_1^{+}$, where
$$
L_1^{+}=\frac d{dx}\ln v_1-D,\quad L_2^{+}=\frac d{dx}\ln \frac{W(v_1,%
\widetilde{v}_1)}{v_1}-D\enspace.
$$
Using expression (\ref{e9}), we obtain
\begin{equation}
\label{LNC}L^{(N)^{+}}=-L^{(N-2)^{+}}(h_N-C_1)\enspace .
\end{equation}
Here, $L^{(N-2)^{+}}$ is an $(N-2)$-order Darboux transformation operator
which transforms the solutions of the Schr\"odinger equation with the
Hamiltonian $h_N$ to the solutions of the same equation with the Hamiltonian
$h_0$. It follows from (\ref{LNC}) that $L^{(N)}=-L^{(N-2)}(h_0-\overline{C}%
_1)$. When $C_1\in \bbbr,$ the latter expression contradicts to (\ref{e8})
and when $C_1\in \bbbc
$ it leads to the lemma's assertion since $\overline{C}_1$ in this case is
as well the zero of the polynomial $P\left( x\right) $. 
$\square $

We can now formulate and prove the main theorem.

{\sc Theorem 1.}{\it \ The action of every nontrivial transformation
operator $L^{(N)}$ is equivalent to the resulting action of any chain of $k$
first order Darboux transformation operators.}

Using Lemmas 2 and 3 and the induction, we can always present an
operator $L^{\left( N\right) }$ in the form $L^{(N)}=L_NL_{N-1}\ldots L_1$
which corresponds to a chain of $N$ first order Darboux transformations. If
we are under the hypothesis of Remark 2, then some of the products of the
operators from this chain are trivial transformation operators. In this case
we have $L^{(N)}=L^{(k)}P(h_0)$, where $L^{(k)}=L_{t+k}L_{t+k-1}\ldots L_t$
with any $t$ and $P(x)$ is some polynomial. The transformation operators $%
L^{(N)}$ and $L^{(k)}$ give the same potential difference $A_N(x)$, and the
theorem is proved. 
$\square $

{\sc Remark 4.}{\it \ This theorem stresses the existence of trivial
transformation operators which can appear in any transformation chain for an
arbitrary initial potential $V_0\left( x\right) $. There exist potentials
(see \cite{FA}) for which the chain of $k\geq 2$ first order operators gives
a trivial transformation operator, while the product of $k-1$ operators is a
nontrivial transformation operator.}

We notice as well that the operator $L^{(k)^{+}}$ realizes the
transformation from the solutions of the Schr\"odinger equation with the
potential $V_N$ to the solutions of equation (\ref{e1}) and consequently can
be used for construction of the operator $L^{(k)^{-1}}$. Moreover, if $%
C_1,\ldots ,C_q$ are different zeros of the polynomial $%
P(h_0)=L^{(k)^{+}}L^{(k)}$, then for the space $T_0$ we can write the
following decomposition:
$$
\begin{array}{c}
T_0=T_{01}\cup \bigcup\nolimits_{i=1}^q
\mathop{\rm span}\left\{ \widetilde{u}_i\right\} , \\ \ker (h_0-C_i)=%
\mathop{\rm span}\left\{ u_i,\widetilde{u}_i\right\} ,\quad i=1,\ldots ,q%
\enspace .
\end{array}
$$
The functions $u_i$ are the transformation functions for the intermediate
transformation operator $L^{(q)}=L_qL_{q-1}\ldots L_1$. A similar
decomposition is valid for the space $T_N$.

\vspace{3mm} \noindent {\bf 3. Potentials with given discrete spectrum on semiaxis}

\vspace{3mm}\noindent We will now apply the Darboux transformation operator
technique to generate and investigate one class of potentials with $N$
discrete energy levels disposed in a desirable manner. To construct such
potentials, one usually uses integral transformation operators. For the
spectral problem on full real axis, these potentials are known as $N$%
-soliton ones (see, e.g., \cite{Newell}). The properties of the $N$-soliton
solutions of the {\sc k}d{\sc v} equation were studied by Wadati and Toda
\cite{WT}. The term '$N$-soliton potential' was introduced by Its and Matveev
\cite{IM}. Sukumar \cite{Suk} and Berezovoi and Pashnev \cite{BP} give a simple
prescription for getting $N$-soliton potentials by means of a chain of
Darboux transformations. Recently (\cite{Jph}) an analytic expressions for
these potentials and for discrete spectrum eigenfunctions have been
obtained.

Following the papers \cite{Suk,BP,tmf,Jph}, we
consider the following solutions of the Schr\"odinger equation with a zero
potential ($h_0=-D^2$) as transformation functions:
\begin{equation}
\label{e10}
\begin{array}{c}
u_k(x)=\cosh \left( a_kx+b_k\right) ,\quad u_{k+1}(x)=\sinh \left(
a_{k+1}x+b_{k+1}\right) , \\
k=1,3,5,\ldots ,
\end{array}
\end{equation}
\begin{equation}
\label{e11}h_0u_i=-a_i^2u_i,\quad 0<a_1<a_2<\ldots \enspace .
\end{equation}
The resulting action of such a chain is equivalent to the action of the $N$%
-order transformation operator (\ref{LN}), and the resulting potential
difference is given by (\ref{e7}) with the replacement $W\left(
u_1,u_2\right) \rightarrow W\left( u_1,\ldots ,u_N\right) $.

For arbitrary $N$, the following can be proved:

{\sc Proposition.} {\it The Wronskian of functions (\ref{e10}) is a
linear combination of hyperbolic cosines:
$$
\begin{array}{c}
W(u_1,...,u_N)= \\
2^{1-N}\sum\limits_{(\varepsilon _1,...,\varepsilon
_N)}^{2^{N-1}}\!\!\!\varepsilon _2\varepsilon _4\cdots \varepsilon
_p\prod\limits_{j>i}^N(\varepsilon _ja_j-\varepsilon _ia_i)\cosh
[\sum\limits_{l=1}^N\varepsilon _l(a_lx+b_l)],
\end{array}
$$
where $\varepsilon _i=\pm 1$; the summation is over all different ordered
sets \newline
$(\varepsilon _1,...,\varepsilon _N)$ consisted of $+1$ and $-1$ (sets $%
(\varepsilon _1,...,\varepsilon _N)$and $(-\varepsilon _1,...,-\varepsilon
_N)$ are considered identical); the index at $\varepsilon _p$ is taken equal
to $N$ or $N-1$ with $N$ being even or odd, respectively.}

The proof can be broken down into two parts: \\ a) The following
formula is proved by the induction method:
$$
W(u_1,...,u_N)=2^{-N}\sum\limits_{\varepsilon _i=\pm 1}\varepsilon _2\varepsilon
_4\cdots \varepsilon _p\prod\limits_{j>i}^N(\varepsilon _ja_j-\varepsilon
_ia_i)\exp [\sum\limits_{l=1}^N\varepsilon _l(a_lx+b_l)];
$$
b) The number of different ordered sets $(\varepsilon _1,\ldots ,\varepsilon _N)$
equals $2^N$. As for any set $(\varepsilon _1,\ldots ,\varepsilon _N)$ there is
a set with an opposite sign $(-\varepsilon _1,\ldots ,-\varepsilon _N)$, we can
detach paired members in the sum, which are packed into the hyperbolic
cosines, and in doing so, we obtain the above formula with the number of
members $2^{N-1}$ and the factor $2^{1-N}$.

The potential calculated according to (\ref{e7}) is regular.

Let $\widetilde{u}_i$ be the second linearly independent solution of
equation (\ref{e11}) with the eigenvalue $E=-a_i^2$ and such that $W\left(
u_i,\widetilde{u}_i\right) =1$. The action of operator (\ref{LN}) to $%
\widetilde{u}_i$ gives the following result:
\begin{equation}
\label{e12}\widetilde{v}_i\left( x\right) =L^{\left( N\right) }\widetilde{u}%
_i=W^{\left( i\right) }\left( u_1,\ldots ,u_N\right) W^{-1}\left( u_1,\ldots
,u_N\right) ,
\end{equation}
where $W^{\left( i\right) }\left( u_1,\ldots ,u_N\right) $ is an $\left(
N-1\right) $-order Wronskian constructed from the functions $u_1,\ldots ,u_N$
except for the function $u_i$. Moreover, the limit
\begin{equation}
\label{e13}\lim \limits_{\psi _E\rightarrow u_i}\left( E+a_i^2\right)
^{-1}L^{\left( N\right) }\psi _E\left( x\right) =v_i\left( x\right)
\end{equation}
exists and gives the second linearly independent solution of the
Schr\"o\-din\-ger equation with the potential $V_N\left( x\right) =A_N\left(
x\right) $ corresponding to the eigenvalue $E=-a_i^2$ and such that $W\left(
v_i,\widetilde{v}_i\right) =1$.

Functions (\ref{e12}) with the transformation functions chosen
according to (\ref{e10}), are square integrable on full real axis.
Parameters $b_i$ realize the isospectral deformation of the potential $V_N$,
and for the sake of simplicity we put all $b_i=0$. In this case, for even $N$%
, the parity of functions (\ref{e12}) coincides with the parity of their
number $i$ and for odd $N$ it is opposite to this parity. Hence, the
potential $V_N$ considered on semiaxis for $N=2k$ and $N=2k+1$ has $k$
discrete energy levels. For even $N$, just $N/2$ functions (\ref{e12}) with
even numbers are the discrete spectrum eigenfunctions of the Hamiltonian $h_N
$, and for odd $N$, we should take $\left( N-1\right) /2$ functions with odd
numbers. The discrete spectrum eigenfunctions normalized to unity have the
form
$$
\varphi _i\left( x\right) =\left[ a_i\prod\limits_{j=1\left( j\neq i\right)
}^N\left| a_i^2-a_j^2\right| \right] ^{1/2}\widetilde{v}_i\left( x\right)
\enspace .
$$
They have the proper values equal to $-a_i^2$.

Let us construct now the Jost solution and the Jost function for the
potential $V_N$. One can find their definition, for instance,
in the first book of the Ref. \cite{invscat}.

Let $f^{\left( 0\right) }\left( k\right) =\exp \left( \I kx\right) $%
, $E=k^2$, be the Jost solution for the initial potential. Then the solution
$\psi ^{\left( 0\right) }=\frac 1k\sin \left( kx\right) $ of the same
equation, regular at $x=0$, is expressed with the help of the Jost solution
as follows:
\begin{equation}
\label{e14}\psi ^{\left( 0\right) }=\frac 1{2k\I }\left[ f^{\left( 0\right)
}\left( k\right) -f^{\left( 0\right) }\left( -k\right) \right] \enspace .
\end{equation}
We need as well the second solution of the same equation corresponding to $%
E=k^2$:
\begin{equation}
\label{e15}\widetilde{\psi }^{\left( 0\right) }=\cos \left( kx\right) =\frac
12\left[ f^{\left( 0\right) }\left( k\right) +f^{\left( 0\right) }\left(
-k\right) \right] \enspace .
\end{equation}
It can readily be seen that the behavior of the functions $L^{\left(
N\right) }\psi ^{\left( 0\right) }$ and $L^{\left( N\right) }\widetilde{\psi
}^{\left( 0\right) }$ as $x\rightarrow 0$ depends on the parity of $N$. For
an even $N$, we have%
$$
L^{\left( N\right) }\psi ^{\left( 0\right) }\cong \left( -1\right)
^{N/2}\prod\limits_{i=1}^{N/2}\left( k^2+a_{2i}^2\right) x,
$$
$$
L^{\left( N\right) }\widetilde{\psi }^{\left( 0\right) }\cong \left(
-1\right) ^{N/2}\prod\limits_{i=1}^{N/2}\left( k^2+a_{2i-1}^2\right)
$$
and for an odd $N$ we obtain%
$$
L^{\left( N\right) }\psi ^{\left( 0\right) }\cong \left( -1\right) ^{\left(
N-1\right) /2}\prod\limits_{i=1}^{\left( N-1\right) /2}\left(
k^2+a_{2i}^2\right) ,
$$
$$
L^{\left( N\right) }\widetilde{\psi }^{\left( 0\right) }\cong \left(
-1\right) ^{\left( N+1\right) /2}\prod\limits_{i=0}^{\left( N-1\right)
/2}\left( k^2+a_{2i+1}^2\right) x\enspace .
$$
It is easily seen from these relations that the solution regular at the
origin (i.e., the solution with the asymptotic $x^{-1}\psi ^{\left( N\right)
}\rightarrow 1$ as $x\rightarrow 0$) has the form
\begin{equation}
\label{e16}\psi ^{\left( N\right) }=\left( -1\right) ^{\left( N+1\right)
/2}\prod\limits_{i=0}^{\left( N-1\right) /2}\left( k^2+a_{2i+1}^2\right)
L^{\left( N\right) }\widetilde{\psi }^{\left( 0\right) }
\end{equation}
for an odd $N$ and
\begin{equation}
\label{e17}\psi ^{\left( N\right) }=\left( -1\right)
^{N/2}\prod\limits_{i=0}^{N/2}\left( k^2+a_{2i}^2\right) L^{\left( N\right)
}\psi ^{\left( 0\right) }
\end{equation}
for an even $N$.

By the same means, the behavior of the function $L^{\left( N\right)
}f^{\left( 0\right) }\left( k\right) $ as $x\rightarrow \infty $,%
$$
L^{\left( N\right) }f^{\left( 0\right) }\left( k\right) \cong
\prod_{j=1}^N\left( -a_j+\I k\right) \exp \left( \I kx\right) ,
$$
permits one to write the Jost solution for the potential $V_N$%
\begin{equation}
\label{e18}f^{\left( N\right) }\left( k\right) =\prod_{j=1}^N\left( -a_j+\I %
k\right) ^{-1}L^{\left( N\right) }f^{\left( 0\right) }\left( k\right)
\enspace .
\end{equation}

With the help of relations (\ref{e14}), (\ref{e15}), and \ref{e18}),
equations (\ref{e16}) and (\ref{e17}) can be rewritten as follows:
\begin{equation}
\label{e19}\psi ^{\left( N\right) }=\frac{\I }{2k}\left[ F^{\left( N\right)
}\left( k\right) f^{\left( N\right) }\left( -k\right) -F^{\left( N\right)
}\left( -k\right) f^{\left( N\right) }\left( k\right) \right] ,
\end{equation}
where
\begin{equation}
\label{e20}F^{\left( N\right) }\left( k\right) =\prod_{j=N/2}^\gamma \frac{k-%
\I a_{2j-1}}{k+\I a_{2j}}
\end{equation}
is the Jost function for the potential $V_N$; $\gamma =1/2$ when $N$ is odd
and $\gamma =1$ when $N$ is even; $a_0=0$. Equation (\ref{e20}) clearly
shows that the potential $V_N$ is of the Bargman type.

\vspace{3mm} \noindent {\bf 4. Examples}

\vspace{3mm}\noindent Consider now simplest particular cases.
For $N=1$ the potential $V_N$ is the well known one soliton
potential
$$
V_1=-2a_1^2\mathop{\rm sech}\nolimits^2\left( a_1x\right) \enspace .
$$
This potential, for the spectral problem on full real axis, has one discrete
energy level and, when being considered on semiaxis, it has no discrete
spectrum at all. Using the transformation operator $L^{\left( 1\right) }$,
we easily obtain its Jost solution:%
$$
f^{\left( 1\right) }\left( k\right) =\frac{a_1\tanh \left( a_1x\right) -\I k%
}{a_1-\I k}\exp \left( \I kx\right) \enspace .
$$
According to formula (\ref{e20}), we obtain the Jost function $F^{\left(
1\right) }\left( k\right) =k\left( k+\I a_1\right) ^{-1}$. The module and
phase of this function, $F^{\left( 1\right) }\left( k\right) =F_1\E ^{-\I %
\delta }$, define the asymptotic of the regular solution of the
Schr\"odinger equation at $x\rightarrow \infty $:%
$$
\psi ^{\left( 1\right) }=\frac{\I }{2k}\left[ \frac k{k+\I a_1}f^{\left(
1\right) }\left( k\right) -\frac{-k}{-k+\I a_1}f^{\left( 1\right) }\left(
-k\right) \right] ,
$$
$$
\psi ^{\left( 1\right) }\rightarrow \frac{F_1}k\sin \left( kx+\delta \right)
\enspace .
$$

It is interesting to note that the choice $b_1\neq 0$ corresponds to
the translation of the origin along the $x$ axis by the value $\Delta
=-b_1/a_1$. Choosing values of $b_1$ and $a_1$, we can give an arbitrary
value to the quantity $\Delta $. Nevertheless, the potential $V_1$ not
necessary has a discrete spectrum. This means that the absence on semiaxis
and the presence on full real axis of a discrete spectrum are due to the
asymptotic behavior of the potential $V_1$.

For $N=2$ we have the two soliton potential%
$$
V_2=\frac{8\left( a_1^2-a_2^2\right) \left[ a_2^2\cosh \left( a_1x\right)
+a_1^2\sinh \left( a_2x\right) \right] }{\left[ \left( a_2+a_1\right) \cosh
\left[ \left( a_2-a_1\right) x\right] +\left( a_2-a_1\right) \cosh \left[
\left( a_2+a_1\right) x\right] \right] ^2}
$$
having a single discrete spectrum level equal to $-a_1^2$ with the wave
function%
$$
\varphi _0=\frac{2\sqrt{a_1\left( a_2^2-a_1^2\right) }\sinh \left(
a_2x\right) }{\left( a_2+a_1\right) \cosh \left[ \left( a_2-a_1\right)
x\right] +\left( a_2-a_1\right) \cosh \left[ \left( a_2+a_1\right) x\right] }%
\enspace .
$$
Its Jost function has the form:%
$$
F^{\left( 2\right) }\left( k\right) =\frac{k-\I a_1}{k+\I a_2}\enspace .
$$

\vspace{3mm}\noindent {\bf 5. Discussion and concluding remarks }

\vspace{3mm}\noindent We have shown that every $N$-order Darboux
transformation operator defined in terms of a general notion of transformation operators may be presented as a chain of $k(\le N)$ usual Darboux transformation operators ( so called a dressing chain \cite {FA}). It is worth stressing that this chain may h
ave ill defined elements. There are two possibilities for such elements. In the first case we may obtain complex-valued potentials for some intermediate Hamiltonians. Corresponding eigenvalue problem can not be treated as Schr\"odinger equation and no qua
ntum mechanical interpretation exists for eigenfunctions. In the second case some intermediate potentials may have poles in the interval for the variable $x$ where the initial Schr\"odinger equation is defined. In this case several spectral problems arise
s. In general, the spectrum of such problems has no common points with the initial problem. Nevertheless, we may obtain solution to these auxiliary spectral problems (if necessary) with the help of the Darboux transformation operator method. Such a situat
ion is studied in detail in \cite{BOSJMPS}. In the case of the absence of ill defined elements in the chain of transformations we have {\it a completely reducible chain}. Conception of reducibility of chains of Darboux transformations with respect to comp
lex-valued intermediate potentials has been introduced by Andrianov et. al. \cite{andr}. With regard to these remarks we state that the study of $N$-order Darboux transformation operators may be reduced to the study of different chains of the first order 
Darboux transformations.

The above statement has been illustrated by the application of a completely reducible chain of transformations to create a potential with arbitrarily disposed discrete spectrum levels for the spectral problem on semiaxis. A closed formula for an $N$-order
 Wronskian has been given. To obtain the potential it is sufficient to take the second logarithmic derivative of this Wronskian. With the help of the $N$-order Darboux transformation operator the Jost solution and the Jost function have been obtained. Our
 analysis shows that the presence of a single discrete level and its absence when the same potential is cosidered on full real axis and on semiaxis is due to the assymptotic behaveour of the potential.

\vspace{3mm}\noindent {\bf Acknowledgments}

\vspace{3mm}\noindent This work is partially supported by RFBR grant No 97-02-16279.

\end{document}